\documentclass[conference]{IEEEtran}
\IEEEoverridecommandlockouts

\usepackage{amsmath,amssymb}
\usepackage{newtxmath}
\usepackage{array}
\usepackage{booktabs}
\usepackage{cite}
\usepackage{graphicx}
\usepackage{microtype}
\usepackage[table]{xcolor}

\newif\ifresultsavailable
\resultsavailabletrue

\title{Construction-Constrained Bernstein Synthesis of Joint Pitch--Width RDL Spiral Inductors for High-$Q$ mmWave FoWLP Applications}

\author{%
\begin{tabular*}{\textwidth}{@{\extracolsep{\fill}}cccc@{}}
\parbox[t]{0.22\textwidth}{\centering
Shuhang Chen\\[0.6ex]
\footnotesize\textit{Department of Electronic Engineering and Information Science}\\
\textit{University of Science and Technology of China}\\
Hefei, Anhui 230027, China\\
chenshuhang@mail.ustc.edu.cn}
&
\parbox[t]{0.22\textwidth}{\centering
Xinmeng Wang\\[0.6ex]
\footnotesize\textit{Department of Electronic Engineering and Information Science}\\
\textit{University of Science and Technology of China}\\
Hefei, Anhui 230027, China\\
xinmengwang@mail.ustc.edu.cn}
&
\parbox[t]{0.22\textwidth}{\centering
Yuchong Sun\\[0.6ex]
\footnotesize\textit{Department of Electronic Engineering and Information Science}\\
\textit{University of Science and Technology of China}\\
Hefei, Anhui 230027, China\\
ustcsyc@mail.ustc.edu.cn}
&
\parbox[t]{0.22\textwidth}{\centering
Qi Zhu\\[0.6ex]
\footnotesize\textit{Department of Electronic Engineering and Information Science}\\
\textit{University of Science and Technology of China}\\
Hefei, Anhui 230027, China\\
zhuqi@ustc.edu.cn}
\end{tabular*}}

\begin{document}
\maketitle

\begin{abstract}
We present a construction-constrained Bernstein framework for joint pitch--width synthesis of planar spiral inductors. A positive cubic Bernstein profile is integrated to prescribe a monotonic radial trajectory, while an independent Bernstein profile controls the conductor width. The nonnegativity, partition-of-unity, and convex-hull properties of the Bernstein basis convert endpoint, winding-direction, and width requirements into algebraic constraints. Embedded in differential evolution (DE), these constraints reject inadmissible candidates before full-wave evaluation, focusing expensive solver calls on geometries that satisfy the construction rules. The method is demonstrated on a two-turn mmWave inductor implemented in the redistribution layer (RDL) of fan-out wafer-level packaging (FoWLP), with the quality factor selected as the optimization objective. Searching the constrained Bernstein design space with DE yields a simulated $\boldsymbol{Q}$ of 34.85 at 30~GHz, compared with 29.66 for a uniform Archimedean baseline under the same stack, footprint, and ports. This corresponds to a 17.5\% improvement while reducing planar copper area by 1.53\%. The optimized response maintains $\boldsymbol{Q}\geq\text{\textbf{30}}$ from 16 to 49~GHz.

\end{abstract}


\section{Introduction}

\begin{figure}[!b]
\centering
\vspace{-6pt}
\includegraphics[width=\columnwidth,trim=13 11 13 11,clip]{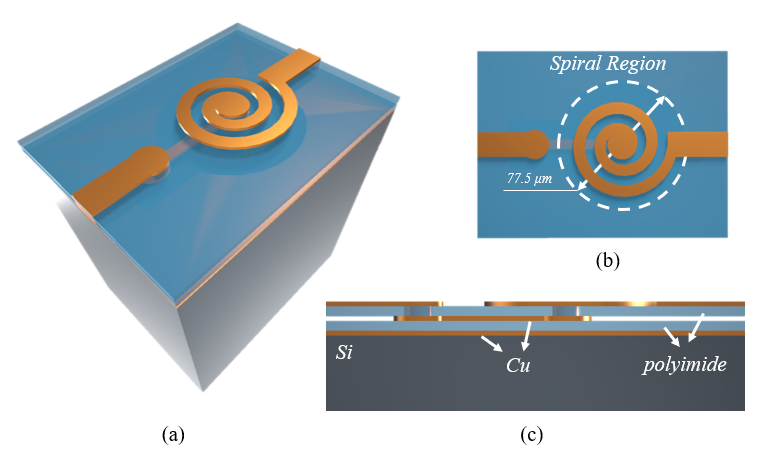}
\vspace{-8pt}
\caption{FoWLP setting used to instantiate the geometry-design problem. (a) Three-dimensional package model. (b) Designated top-layer spiral region within the ground opening. (c) Fixed Si/Cu/polyimide stack and multilayer interconnect.}
\label{fig:model_stack}
\end{figure}

Planar spiral inductors (PSIs) are essential passive components in matching networks \cite{shaeffer1997lna}, resonators \cite{chernobryvko2024resonator}, oscillators \cite{cao2025classf23}, RF filters \cite{he2005lcfilter}, millimeter-wave power amplifiers \cite{hsieh2025cui}, and wireless power transfer \cite{sample2011wpt,chen2024nonlinearwpt}. At mmWave frequencies, low-loss, high-$Q$ implementations are particularly valuable for RF front ends, as recently demonstrated in Ka/V-band heterogeneous-integrated LNAs \cite{sun2025siinterposer} and mmWave/sub-THz RF-interposer circuits \cite{wallner2026rfinterposer}. PSIs have therefore been realized across diverse IC and advanced-packaging platforms, including CMOS \cite{shaeffer1997lna}, Si BiCMOS \cite{tayenjam2019variablepitch}, FoWLP \cite{murugesan2021fowlp}, and FoPLP \cite{niu2021foplp}. Once the fabrication process, material stack, design rules, and available footprint are fixed, geometry becomes the principal remaining design freedom for optimizing inductance, loss, parasitic capacitance, self-resonance, bandwidth, coupling, or their tradeoffs. Prior work progressed from variable-width spirals \cite{chen2022variablewidth} and variable-pitch spirals \cite{tayenjam2019variablepitch} to simultaneous width--spacing variation. Square spirals with turn-wise arithmetic width and geometric spacing progressions \cite{shen2011nonuniform} and tapered spirals with turn-wise width--spacing variation at fixed pitch \cite{sathyasree2018proximity} demonstrated the benefit of coordinating both variables. Continuous width modulation extends beyond turn-wise tapering but does not independently control pitch \cite{canada2025continuous}, whereas deformable mmWave spirals rely on predefined deformation dimensions and segment-wise width variation \cite{wei2023deformable}. These limitations motivate a representation that can continuously shape both the spiral trajectory and conductor width within a prescribed electromagnetic environment. Fig.~\ref{fig:model_stack} instantiates this design setting for the FoWLP platform considered here.

In Fig.~\ref{fig:model_stack}, the package stack, ground opening, multilayer interconnect, and ports establish the surrounding electromagnetic environment, while the planar trace occupies the designated top-layer region. Originating in approximation theory and widely used in geometric design, the Bernstein basis is nonnegative, forms a partition of unity, and has the convex-hull property \cite{farouki2012bernstein}. Independent Bernstein profiles can thus describe the trace trajectory and conductor width without imposing a fixed linear, arithmetic, or geometric law. These properties turn radial endpoints, monotonic winding, and width bounds into coefficient constraints, rejecting inadmissible candidates before full-wave evaluation. The explicit Bernstein profiles also allow the physical strip edges to be reconstructed directly, making clearance and self-intersection straightforward to check.

This paper develops a construction-constrained Bernstein parameterization for joint pitch--width synthesis. The method is applied to a high-$Q$ mmWave inductor implemented in the redistribution layer (RDL) of fan-out wafer-level packaging (FoWLP), whose thick copper and low-loss polymers can reduce loss relative to on-chip implementations \cite{wojnowski2012ewlb,murugesan2021fowlp}. Its compact dimension is independent of boundary discretization and is directly compatible with full-wave optimization. Under identical stack, footprint, ports, and solver settings, the optimized spiral reaches a simulated $Q$ of 34.85 at 30~GHz, a 17.5\% improvement over the 29.66 uniform Archimedean baseline, while using 1.53\% less planar copper area. The claims are limited to intrinsic full-wave simulation.

\section{Construction-Constrained Bernstein Geometry}

\begin{figure}[!h]
\centering
\includegraphics[width=\columnwidth]{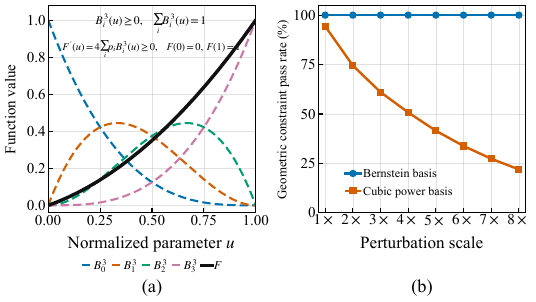}
\vspace{-16pt}
\caption{Bernstein construction and coefficient-space sensitivity. (a) Cubic basis functions and a normalized cumulative-radius function. (b) Geometric constraint pass rate under scaled local perturbations; each point aggregates 30,000 candidates from three seeds. Passing requires radial endpoints, radial monotonicity, and width bounds; clearance and self-intersection are excluded and checked separately.}
\label{fig:bernstein_basis}
\end{figure}

To show why the Bernstein basis is well suited for parameterizing spiral geometries with continuously varying pitch and conductor width, we derive the radial and width profiles from the cubic basis shown in Fig.~\ref{fig:bernstein_basis}(a).

\begin{figure}[!b]
\centering
\vspace{-6pt}
\includegraphics[width=\columnwidth,trim=3 3 3 3,clip]{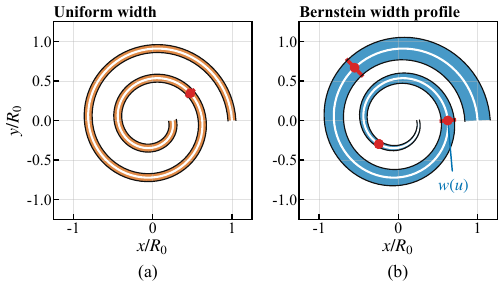}
\vspace{-8pt}
\caption{Normalized geometry. (a) Uniform-width Archimedean baseline. (b) Bernstein-controlled width profile. Red cross-sections mark local widths and white curves denote centerlines.}
\label{fig:bernstein_spiral}
\end{figure}

Let $u\in[0,1]$ be the normalized centerline parameter running from the outer to the inner endpoint, and let $\Theta$ denote the total winding angle, with $\Theta=4\pi$ for the two-turn design. For $i\in\{0,1,2,3\}$, the cubic Bernstein basis function is $B_i^3(u)=\binom{3}{i}u^i(1-u)^{3-i}$. On $u\in[0,1]$, the basis is nonnegative and forms a partition of unity: $B_i^3(u)\geq0$ and $\sum_{i=0}^{3}B_i^3(u)=1$. We first exploit nonnegativity by choosing normalized radial-contraction weights $p_i>0$ with $\sum_{i=0}^{3}p_i=1$. Their cumulative profile is
\begingroup
\setlength{\abovedisplayskip}{3pt}
\setlength{\abovedisplayshortskip}{2pt}
\setlength{\belowdisplayskip}{3pt}
\setlength{\belowdisplayshortskip}{3pt}
\begin{equation}
F(u)=4\int_0^u\sum_{i=0}^{3}p_iB_i^3(v)\,\mathrm{d}v,
\label{eq:cumulative_profile}
\end{equation}
\endgroup
where $v$ is the integration variable. Although the construction extends to degree $n$ with $n+1$ coefficients and normalization factor $n+1$, cubic profiles limit optimization dimension and discourage variations finer than fabrication or numerical resolution. Let $R_0$ be the outer centerline radius and let $\alpha\in(0,1)$ be the inner-to-outer centerline-radius ratio. The polar radius $r(u)$ and Cartesian centerline $\mathbf{c}(u)$ are
\begin{equation}
r(u)=R_0[1-(1-\alpha)F(u)],\qquad
\mathbf{c}(u)=r(u)
\begin{bmatrix}\cos\Theta u\\ \sin\Theta u\end{bmatrix}.
\label{eq:centerline}
\end{equation}
Here and below, a prime denotes differentiation with respect to $u$. Nonnegativity of the Bernstein basis, together with $p_i>0$, gives $F'(u)=4\sum_{i=0}^{3}p_iB_i^3(u)>0$ and therefore $r'(u)<0$, so the winding cannot reverse radially. In addition, each cubic basis function satisfies $\int_0^1B_i^3(v)\,\mathrm{d}v=1/4$; combined with $\sum_i p_i=1$, this gives $F(0)=0$ and $F(1)=1$. Consequently, $r(0)=R_0$ and $r(1)=\alpha R_0$, so both radial endpoints are exact.

Let $\beta_i$ be dimensionless width coefficients. The normalized full-width profile $w(u)$ and physical full width $W(u)$ are
\begin{equation}
w(u)=\sum_{i=0}^{3}\beta_iB_i^3(u),\qquad W(u)=R_0w(u).
\label{eq:width}
\end{equation}
Writing $\mathbf{c}(u)=[c_x(u),c_y(u)]^\mathsf{T}$, its unit normal is $\mathbf{n}(u)=[-c_y'(u),c_x'(u)]^\mathsf{T}/\|\mathbf{c}'(u)\|_2$, where $\|\cdot\|_2$ denotes the Euclidean norm. The two physical strip-boundary curves $\boldsymbol{\Gamma}_{\pm}(u)$ are
\begin{equation}
\boldsymbol{\Gamma}_{\pm}(u)=\mathbf{c}(u)\pm\frac{W(u)}{2}\mathbf{n}(u).
\label{eq:edges}
\end{equation}
Nonnegativity and partition of unity now give the convex-hull property: $w(u)$ is a convex combination of the width coefficients $\beta_i$. It follows directly that $R_0\cdot\min_i\beta_i\leq W(u)\leq R_0\cdot\max_i\beta_i$, so coefficient bounds enforce the full-width limits over the entire curve rather than only at sampled locations. For fixed $\alpha$ and $\Theta$, the four normalized $p_i$ contain three independent radial-profile degrees of freedom; together with the four $\beta_i$, they give seven design variables independent of boundary discretization.

Fig.~\ref{fig:bernstein_basis}(b) quantifies the resulting improvement in geometric constraint pass rate. As the coefficient-perturbation scale increases from $1\times$ to $8\times$, the Bernstein basis retains a 100\% pass rate without curve repair, whereas an endpoint-constrained cubic power basis falls from 94.37\% to 21.94\%. The test covers endpoints, radial monotonicity, and width bounds.

\section{FoWLP Case Study, Optimization, and Results}

The Bernstein construction is applied to the two-turn mmWave FoWLP redistribution-layer (RDL) inductor shown in Fig.~\ref{fig:model_stack}.

\begin{figure*}[!t]
\centering
\begin{minipage}[t]{0.56\textwidth}
\centering
\includegraphics[height=0.245\textheight,trim=10 7 10 4,clip]{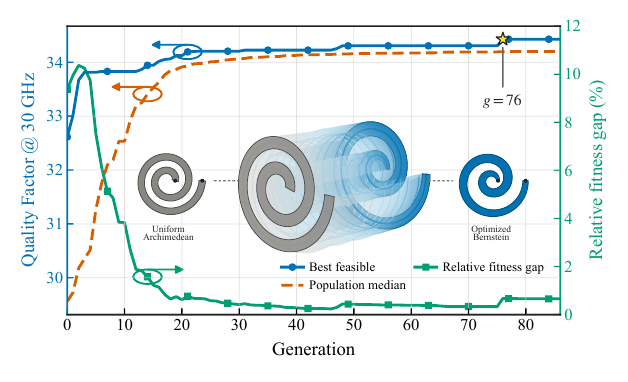}
\par\vspace{-8pt}{\fontsize{9}{10}\selectfont (a)}
\end{minipage}\hfill
\begin{minipage}[t]{0.42\textwidth}
\centering
\includegraphics[height=0.245\textheight,trim=5 0 5 0,clip]{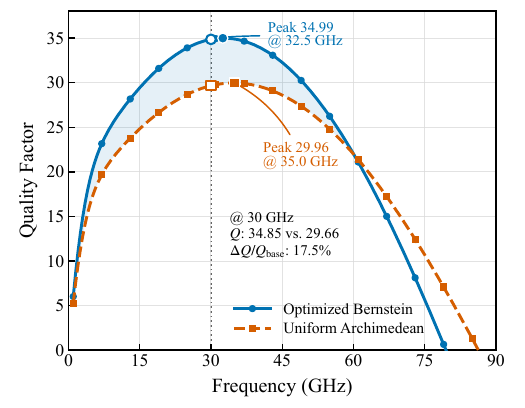}
\par\vspace{-8pt}{\fontsize{9}{10}\selectfont (b)}
\end{minipage}
\vspace{2pt}
\caption{Optimization and response. (a) Constrained-SHADE histories at 30~GHz. The right-axis relative fitness gap, $100(Q_{\mathrm{best}}-Q_{\mathrm{median}})/Q_{\mathrm{best}}$, approaches zero as the population converges. The inset connects recorded uniform-to-Bernstein incumbents; translucent surfaces are interpolated only for visualization. The final champion appears at generation 76. (b) Intrinsic $Q$ sweep: 34.85 versus 29.66 at 30~GHz, a 17.5\% improvement.}
\label{fig:optimization_and_geometry_evolution}
\end{figure*}

Following the material stack in \cite{murugesan2021fowlp}, the model uses a 3-$\mu$m top copper trace above a 6-$\mu$m polyimide layer with $\epsilon_r=3.15$ and $\tan\delta=0.005$. The outer and inner centerline radii are fixed at 62.5 and 12.5~$\mu$m, respectively; conductor width is limited to 5--20~$\mu$m and minimum edge spacing to 5~$\mu$m. The spiral lies on the top copper layer, a fixed middle-layer underpass connects its inner endpoint to the second pad, and the bottom layer is the reference ground.

The uniform Archimedean baseline uses $p_i=0.25$ and $\beta_i=0.192$, corresponding to a 12-$\mu$m conductor width. It shares the footprint, material stack, underpass, ports, and solver settings with the optimized design, so only the centerline contraction and conductor-width profiles vary. The optimized coefficients are $(p_0,p_1,p_2,p_3)=(0.0002,0.9044,0.0953,0.0001)$ and $(\beta_0,\beta_1,\beta_2,\beta_3)=(0.1574,0.2701,0.3171,0.0805)$. Let $f$ denote frequency and $Y_{11}(f)$ the $(1,1)$ entry of the simulated two-port admittance matrix. The intrinsic quality factor $Q(f)$ and inductance $L(f)$ are extracted as
\begin{equation}
Q(f)=-\frac{\operatorname{Im}Y_{11}(f)}{\operatorname{Re}Y_{11}(f)},\qquad
L(f)=\frac{\operatorname{Im}\{1/Y_{11}(f)\}}{2\pi f}.
\label{eq:lq}
\end{equation}
Only frequencies with $\operatorname{Re}Y_{11}(f)>0$ are retained.

Constrained success-history-based adaptive differential evolution (SHADE) \cite{tanabe2013shade} generates trial coefficient vectors to maximize $Q(30\,\mathrm{GHz})$, with each objective value obtained from an HFSS full-wave evaluation of the corresponding admissible geometry. The search is subject to a $\pm15\%$ inductance tolerance about 450~pH and $\operatorname{Re}Y_{11}(30\,\mathrm{GHz})>0$. Before each HFSS call, the radial weights and width coefficients are projected onto the algebraic constraints established in Section~II, followed by minimum-clearance and self-intersection checks on the generated strip. Thus, inadmissible trials are rejected before the expensive full-wave objective evaluation.

\begin{table*}[!t]
\caption{Comparison of Nonuniform Spiral Geometry Representations}
\label{tab:literature}
\centering
\setlength{\tabcolsep}{3.0pt}
\scriptsize
\renewcommand{\arraystretch}{0.86}
\begin{tabular}{@{}
>{\raggedright\arraybackslash}m{0.12\textwidth}
>{\raggedright\arraybackslash}m{0.12\textwidth}
>{\raggedright\arraybackslash}m{0.15\textwidth}
>{\raggedright\arraybackslash}m{0.205\textwidth}
>{\raggedright\arraybackslash}m{0.225\textwidth}
>{\raggedright\arraybackslash}m{0.09\textwidth}@{}}
\toprule
Work & Width profile & Turn-separation profile & Geometry / process-rule handling & Comparison case & Relative $Q$ improvement (\%) \\
\midrule
Shen \emph{et al.} \cite{shen2011nonuniform} & Discrete by turn & Discrete by turn & Prescribed arithmetic-width/geometric-spacing progressions & Versus uniform width/spacing; 3.5-turn square; simulation & 42.86\% \\
Sathyasree \emph{et al.} \cite{sathyasree2018proximity} & Discrete by turn & Discrete by turn & Coupled taper at fixed centerline pitch & Versus untapered; 5.5-turn octagonal; measurement & 43\% \\
Ca\~nada \emph{et al.} \cite{canada2025continuous} & Continuous linear & Fixed & Prescribed width range & Versus constant-wide of similar $L$; circular PCB; measurement & 9.1--9.4\% \\
\rowcolor{blue!8}\textbf{This work} & \textbf{Continuous Bernstein} & \textbf{Continuous centerline pitch} & \textbf{Coefficient-bounded widths; edge clearance screened} & \textbf{Versus uniform Archimedean; two-turn FoWLP; simulation at 30~GHz} & \textbf{17.5\%} \\
\bottomrule
\end{tabular}
\vspace{3pt}

\parbox{0.985\textwidth}{\footnotesize Relative improvements use the baseline within each study.}
\end{table*}

Fig.~\ref{fig:optimization_and_geometry_evolution}(a) shows convergence and the corresponding geometry evolution through generation 86. The final champion first appears at generation 76, with search-point values of $Q=34.43$ and $L=389.58$~pH. The frequency response in Fig.~\ref{fig:optimization_and_geometry_evolution}(b) is subsequently evaluated for this design.

The geometry inset in Fig.~\ref{fig:optimization_and_geometry_evolution}(a) links convergence to the evolving spiral. At first glance, the uniform Archimedean and final Bernstein geometries remain strikingly similar: both retain two turns and the same footprint, with no added layer or auxiliary structure. Quantitatively, the actual width spans 5.03--15.64~$\mu$m, the minimum edge spacing is 6.53~$\mu$m, centerline length decreases from 474.42 to 424.16~$\mu$m, and copper area decreases from 5693.01 to 5606.02~$\mu$m$^2$ (1.53\%). Yet this visually modest redistribution raises $Q$ at 30~GHz from 29.66 to 34.85, a 17.5\% improvement, while $L$ rises from 349.79 to 391.14~pH. Thus, a comparatively subtle planar reshaping produces a pronounced electrical response.

As shown in Fig.~\ref{fig:optimization_and_geometry_evolution}(b), the optimized design peaks at $Q=34.99$ near 32.5~GHz and maintains $Q\geq30$ from 16 to 49~GHz. Its estimated self-resonant frequency decreases moderately from 86.3 to 79.5~GHz, exposing the bandwidth tradeoff accompanying the $Q$ improvement.

Table~\ref{tab:literature} separates variable geometry from continuous geometry: the former includes turn-wise discrete changes, whereas the latter varies along the trace. Unlike width-only modulation and discrete or coupled turn-wise laws, the Bernstein representation independently controls continuous centerline-pitch and width profiles. Its coefficient bounds are set directly from the process width limits, and the reconstructed edges are screened for clearance. In the optimized layout, the 5.03--15.64~$\mu$m width range and 6.53-$\mu$m minimum spacing satisfy the prescribed 5-$\mu$m limits. Only the top-layer mask geometry changes; the material stack, layer count, and fabrication flow remain unchanged.

To reduce the cost of the DE search, exploratory HFSS evaluations use fewer convergence passes and a coarser mesh; the final champion and baseline are recomputed with tighter convergence settings for a more accurate comparison.

\section{Conclusion}

We introduced a construction-constrained Bernstein parameterization for joint pitch--width control and demonstrated it on a two-turn fan-out RDL spiral. By embedding radial, width, clearance, and winding requirements in the coefficient space, the method generates physically admissible geometries before full-wave evaluation. The optimized design reaches $Q=34.85$ at 30~GHz, 17.5\% above the uniform Archimedean baseline, while reducing planar copper area by 1.53\%; it maintains $Q\geq30$ from 16 to 49~GHz.

More broadly, the Bernstein representation is not tied to a particular material stack, frequency band, layer count, or performance metric. Its coefficient bounds can be redefined for the design rules of different processes, while additional profiles can describe multilayer traces, interlayer transitions, and TSV-enabled three-dimensional coils with spatially varying conductor dimensions and spacing. The same construction principle can therefore extend beyond RDL inductors to planar coils for wireless power transfer, spiral resonators, nonuniform transmission lines (NTLs), and related electromagnetic structures in which continuously shaped trajectories and conductor widths provide useful design freedom. Because the resulting layouts satisfy explicit geometric and process constraints, they are directly suitable for mask generation and fabrication; with process-calibrated material parameters and consistent reference planes, the full-wave workflow also provides a practical route to close simulation--measurement agreement. This establishes Bernstein synthesis as a transferable geometry-optimization framework rather than a design procedure specific to the present FoWLP example.

\bibliographystyle{IEEEtran}
\begingroup
\renewcommand{\baselinestretch}{0.98}\selectfont
\bibliography{references}
\endgroup
\end{document}